\documentclass[article,12pt]
{elsarticle}

\usepackage{amssymb}
\usepackage{graphicx}
\usepackage{epstopdf}

\renewcommand{\part}[1]{\left(#1\right)}  
\newcommand{\ket}[1]{\vert {#1} \rangle}  

\newcommand{\nl}{\par\noindent}   

\newtheorem{theorem}{Theorem}[section]
\newtheorem{lemma}{Lemma}[section]

\newtheorem{definition}{Definition}[section]

{\begin{proof}[#1]%
\renewcommand{\qed}{\rule{0mm}{1mm}\hfill{\footnotesize\textsc{q.e.d.}\
(#2)}}}
{\end{proof}
}

\begin{document}

\begin{frontmatter}



\title{Quantum circuit optimization for unitary operators over non-adjacent qudits}


\author{Giuseppe Sergioli}

\address{University of Cagliari,
    Via Is Mirrionis 1, I-09123 Cagliari, Italy.
\\ giuseppe.sergioli@gmail.com}

\begin{abstract}
Within the general context of the architecture in quantum computer design, this paper aims is to provide a general strategy to obtain a block-matrix representation of quantum gates applied to qubits placed in arbitrary positions over an arbitrary dimensional input state. The model is also extended to the framework of quantum computation with qudits. An application in the context of the quantum computational logic is provided.

\end{abstract}


\end{frontmatter}


\section{Introduction}
\label{Intro}

The topic related to the architecture in quantum computer design plays a crucial role for the realization of advanced technologies in quantum computation \cite{Linke2017}. The standard abstract model of quantum computation assumes that interactions between arbitrary (i.e. non adjacent) pairs (or $n$-tuples) of qubits (or qudits) are available. However, physical architectures conveniently use particular constraints on the qubits distribution based on the nearest-neighbor couplings \cite{K,Linke2017}. In principle, these constraints have not incidence in the possibility to perform arbitrary computations, because the $SWAP$ operations can be suitably used; anyway, the use of the $SWAP$ operations is not free of any computational cost. 
On this basis, recent topics related to efficient quantum computing between remote qubits in nearest-neighbor architectures - such as the linear neighbor architectures LNN \cite{K} - are active and important areas of research, also devoted to physical implementations \cite{FHH,TKO}; as an example, the linear neighbor architectures LNN \cite{K}, offer an appropriate approximate method to approach to physical problems regarding trapped ions \cite{HHRB}, liquid nuclear magnetic resonance \cite{LSB} and the original Kane model \cite{Ka}.

Anyway, translating an arbitrary circuit to the LNN architecture is useful only for a restricted class of physical problems and requires a linear increase of the quantum computational cost of $O(n)$ (where $n$ is number of involved qubits \cite{CMS}). In addition, from a more theoretical point of view, the architecture in quantum computer design plays also a crucial role in the very general problem regarding the classical simulation of quantum circuits. As focused by Jozsa and Miyake \cite{JM}, the capability of a classical computer to efficiently simulate a quantum circuit is strictly related to the ``distance" between the qubits on which a given quantum gate operates, i.e. the number of the $SWAP$ operators necessary to simulate the circuit.  The problematic aspect of that scenario drastically increases in case of multi-qubits systems that represents a further ``complication" in the architecture of a quantum computer \cite{MVBM, ZLDLL} and possible implementations \cite{BTP, CWHR}.

At the ground of the arguments described above, it is possible to notice that, from a purely theoretical viewpoint, a systematic investigation aimed to formally simplify the representation of an arbitary unitary operator applied to arbitrary qubits is actually missing in the standard litterature on theoretical quantum computer science.\footnote{As an example, an introductory result was developed by Wilmott \cite{W} that showed how to represent an arbitrary $SWAP$ gate between two qudits by involving $C-Not$ gates only and by following combinatorial considerations. In this paper we follow a different approach, providing a block-matrix  representation of arbitrary unitary operators without involving the composition of other control gates.} The aim of this work is basically devoted to fill this gap. The first purpose is to provide a simple block-matrix representation of arbitrary binary operators applied to two qubits arbitrary placed within a quantum circuit, without involving multiple composition of $SWAP$ gates. Afterwards, the paper describes a general method to extend this result to arbitrary $n$-ary operators, also in the framework of quantum computation with qudits. The computational benefits of this representation in the architecture of a quantum computer are not detailed in this work; however,  
an immediate benefit of this representation can be found in the context of the simulation of quantum circuits trough programming language. All the computational problems described above, especially in case of multi-qubits systems (or multi-qudits), are naturally extended to the context of programming languages that aim to simulate a quantum circuit with a classical computer. Several attempts in writing an arbitrary quantum circuit are given by using very efficient software (such as Wolfram Mathematica) and many tentatives to obtain suitable representations of quantum circuits by classical computers are actually in-progress \cite{Gerdt1,Gerdt2,Gerdt3}. Basing on this very general setting, to get a block-matrix representation of quantum gates could suggest many different solutions to provide the double advantage to simplify the writing of an arbitrary quantum circuit with a given programming language and to reduce the running time of the software.

Furthermore, and from a totally different perspective, this representation turns out to be particularly beneficial also from a theoretical viewpoint. Indeed, a further advantage of this block-matrix representation can be achieved in the context of the quantum computational logic (QCL) \cite{DGG}. The standard QCL involves in the language only one target gates; the representation achieved along this paper, allows to provide a suitable generalization of the QCL where the language is expanded by involving also non one target gates. The last part of this work is devoted to provide a detailed insight of this idea.

\bigskip

The paper is organized as follows: in Section 2 we briefly summarize the standard representation of unary gates; in Section 3 a block-matrix representation of the $SWAP$ gate applied to arbitrary qubits is provided; in Section 4 we give a block-matrix representation of an arbitrary binary gate while the Section 5 is devoted to show the block-matrix representation of the square root of $SWAP$ gate, as an example. In Section 6 we describe a general method to generalize the representation to arbitrary $n$-ary quantum gates or sets of quantum gates. In Section 7 we extend the previous results to the more general framework of qudits.
Section 8 is devoted to show an application of the block-matrix representation in the context of the quantum computational logic. 
Some brief concluding remarks and possible further developments close the paper. 

\section{The standard representation of unary gates}

The input of a quantum circuit is given by a composition of qubits that is mathematically represented by the tensor product operation. Hence, given $k$ qubits $\ket {x_1},\ket {x_2},\cdots,\ket {x_k}\in\mathbb C^2$ the input state given by an ensemble of $k$ qubits is given by $\ket {x_1}\otimes\ket {x_2}\otimes\cdots\otimes\ket {x_k}$ (that, for short, we call quantum register - or \emph{quregister} - and we indicate by $\ket {x_1,x_2,\cdots,x_k}\in\otimes^k\mathbb C^2$).\footnote{From now on, let us assume that any qubit is written in the canonical basis $\mathcal B=\{\ket 0=\left( \begin{array}{cc}
					 1 \\    
             0  \\ 
						\end{array}\right), \ket 1=\left( \begin{array}{cc}
					 0 \\    
             1  \\ 
						\end{array}\right)\}$ .} Let us remark that, given the non-commutativity of the tensor product, the sequence in which any qubit appears in the state is not negligible; in other words, $\ket{x_1,x_2}$ and $\ket{x_2,x_1}$ generally represent two different states.

A quantum circuit is represented by the evolution of the input quregister under the application of some unitary quantum logical gates \cite{H, NC}. Obviously, it is often the case where a unary quantum gate $U^{(1)}$ is applied to only one qubit $\ket{x_i}$ of the input quregister $\ket x=\ket{x_1,\cdots,x_i,\cdots,x_k}$.


In this case it is necessary to 
 extend the dimension of the quantum gate $U^{(1)}$ to the dimension $k$ of the input state in such a way that $U^{(1)}$ acts on $\ket{x_i}$ only and leaves all the other qubits of the input quregister unchanged. In this case the extension of $U^{(1)}$ into $U^{(k)}$ is simple to achieve and its expression assumes the form: $$U^{(k)}=I^{(i-1)}\otimes U^{(1)}\otimes I^{(k-i)}.$$ Indeed, it is straightforward to see that 
\begin{eqnarray*}
& U^{(k)}\ket x=I^{(i-1)}\ket{x_1,\cdots,x_{i-1}}\otimes U^{(1)}\ket {x_i}\otimes I^{(k-i)}\ket{x_{i+1},\cdots,x_{k}}=\\
&=\ket{x_1,\cdots,x_{i-1}}\otimes U^{(1)}\ket {x_i}\otimes \ket{x_{i+1},\cdots,x_{k}}.
\end{eqnarray*}

In a more general case, if a $n$-ary quantum gate $U^{(n)}$ has to be applied to $\ket{x_{m+1},x_{m+2},\cdots,x_{m+n}}$ in a $k$-dimensional circuit (with $n\leq k$), then the extension of $U^{(n)}$ to the dimension $k$ of the input state $\ket x=\ket{x_1,\cdots,x_m,\cdots,x_{m+n},\cdots,x_k}$, is trivially given by:

\begin{eqnarray}\label{uunitary} U^{(k)}=I^{(m-1)}\otimes U^{(n)}\otimes I^{(k-n-(m-1))}.
\end{eqnarray}

Anyway, this representation is simply achieved because the quantum gate $U^{(n)}$ has to be applied to $n$ ``adjacent" qubits of the input quregister $\ket x$ but this represents only a particular computational situation. Indeed, in a general case, where $U^{(n)}$ has to be applied to $n$ not adjacent qubits of the $k$-dimensional input state, then some suitable strategy becomes necessary and the extension of $U^{(n)}$ to the dimension $k$ is no longer straightforward. Even if this problem is extremely common in the standard theory of quantum computation, a very synthetic expression of the Eq. (\ref{uunitary}) in the general case of non adjacent qubits is actually missing.
In the following we provide a strategy to get a block-matrix representation of an arbitrary $n$-ary operator applied to $n$ arbitrary qubits of a $k$-dimensional input state. In order to do this, we prelimary need to make some algebraic consideration of the $SWAP$ gate.


\section{The $SWAP$ gate}
Let us consider the two-qubits state $\ket{x}\otimes \ket y$. The unitary operator able to switch this state is the well known $SWAP$ operator \cite{NC} that will plays a crucial role in the rest of the paper. 
\begin{definition}{SWAP gate}\label{swap}

Let $\ket x$ and $\ket y$ unitary vectors belonging to the Hilbert space $\mathbb C^2$. The $SWAP$ gate is defined as:
$$SWAP (\ket x \otimes \ket y)=\ket y \otimes \ket x.$$

\end{definition}
It is easy to check that the matrix form of $SWAP$ is given by: 

\begin{eqnarray*}
SWAP= \left[ \begin{array}{cccc}
					 1 & 0  & 0 & 0 \\    
					 0 & 0  & 1 & 0 \\ 
             0 & 1  & 0 & 0 \\ 
             0 & 0  & 0 & 1 \\ 
						\end{array}\right].
\end{eqnarray*} 

Let us consider the projectors operators:
 $P_0=\ket 0 \langle 0| =\left[ \begin{array}{cc}
					 1 & 0  \\    
					 0 & 0  \\ 
            
						\end{array}\right]$ and $P_1=\ket 1 \langle 1|=\left[ \begin{array}{cc}
					 0 & 0  \\    
					 0 & 1  \\ 
            
						\end{array}\right]$ and the Ladder operators \cite{F}:

 $L_0=\ket 0 \langle 1| =\left[ \begin{array}{cc}
					 0 & 1  \\    
					 0 & 0  \\ 
            
						\end{array}\right]$ and  $L_1=\ket 1  \langle 0| =\left[ \begin{array}{cc}
					 0 & 0  \\    
					 1 & 0  \\ 
            
						\end{array}\right].$ 

These operators allow to provide the following block-matrix representation of $SWAP$ \begin{eqnarray}\label{blockswap}
SWAP=\left[ \begin{array}{c|c}
					P_0 & L_1 \\ \hline     
					 L_0 & P_1 \\ 
						\end{array}\right],
\end{eqnarray}
that turns out to be useful in the rest of the paper.


In a more general case, the input of a circuit can involve more than two qubits (for istance $\ket {x_1, \cdots x_m, \cdots x_{m+n}, \cdots x_k}$); by resorting to Eq.(\ref{uunitary}) it is easy to swap two adjacent qubits belonging to an arbitrary $k$-dimensional input state, but in a general scenario it could be required to perform an arbitrary $SWAP_{(k;m,m+n)}$ gate such that $$SWAP_{(k;m,m+n)}\ket {x_1, \cdots, x_m, \cdots, x_{m+n}, \cdots x_k}=\ket {x_1, \cdots, x_{m+n}, \cdots, x_m, \cdots, x_k},$$ that is a $SWAP$ between two non-consecutive qubits. Simply speaking, $SWAP_{(k;m,m+n)}$ represents a $SWAP$ gate between the $m$-th and the $(m+n)$-th qubits, in a $k$-input qubits circuit.
Obviously, the $SWAP_{(k;m,m+n)}$ is achievable by compositions of shifted two-qubit $SWAP$ gates. This section is devoted to provide a block-matrix representation \cite{B} of an arbitrary $SWAP_{(k;m,m+n)}$ gate. 


First, let us consider the special case where we apply the $SWAP$ gate between the first and the 
last qubits of a $n$-dimensional input state. We introduce the following Lemma.
\begin{lemma}\label{swap}
Let $\ket{x_1},\cdots, \ket{x_n}\in\mathbb C^2$ and let $P_0^{(n)}=I^{(n-1)}\otimes P_0$ (with $I^{(n-1)}$ $(n-1)$-dimensional identity matrix); similarly for $P_1^{(n)}, L_0^{(n)}$ and $L_1^{(n)}.$ The block-matrix representation of $SWAP_{(n;1,n)}$ is given by:
$$SWAP_{(n;1,n)}=\left[ \begin{array}{c|c}
					P_0^{(n-1)} & L_1^{(n-1)} \\ \hline     
					 L_0^{(n-1)} & P_1^{(n-1)} \\ 
						\end{array}\right].$$
\end{lemma}

Proof:

\bigskip

First, we prove that:

 $SWAP_{(n;1,n)}\ket{x_1,\cdots, x_n}=\ket{x_n,x_2,\cdots, x_{n-1},x_1}$ and afterwards we check the unitarity of
 $SWAP_{(n;1,n)}$.

First, let us prove that $
SWAP_{(n;1,n)}$ provides a swap between the first and the last qubits of a $n$-dimensional input state.

Let $\ket{x} = \ket{x_1, \ldots, x_n}$ be a basis vectors in $\otimes^n{\mathbb{C}^2}$ and let be 

$\ket{x_1} = \left( \begin{array}{c}
   x_{1_a} \\
   x_{1_b} 
 \end{array} \right)~~~and~~~
 \ket{x_n}= \left( \begin{array}{c}
   x_{n_a} \\
   x_{n_b} 
 \end{array} \right)$.
 Using the standard properties of the product of the block matrices, we have that:
\begin{eqnarray*}
& & SWAP_{(n;1,n)}\ket{x_1\cdots x_n}= \\
&=& \left[ 
	\begin{array}{c|c}
	    {P_0^{(n-1)}} & {L_1^{(n-1)}} \\ \hline     
	    {L_0^{(n-1)}} & {P_1^{(n-1)}}	 \\ 
	   \end{array} \right] \cdot 
  \left( \begin{array}{c}
   x_{1_a} \\
   x_{1_b} 
  \end{array} \right) \otimes 
 \ket{x_2, \ldots, x_{n-1}} \otimes 
 \left( \begin{array}{c}
   x_{n_a} \\
   x_{n_b} 
 \end{array} \right)= \\ 
& = & \left[ \begin{array}{c|c}
	    {P_0^{(n-1)}} & {L_1^{(n-1)}} \\ \hline     
	    {L_0^{(n-1)}} & {P_1^{(n-1)}}	 \\ 
	   \end{array} \right] \cdot 
	   \left(
  \begin{array}{c}
         x_{1_a}~\ket{x_2, \ldots, x_{n-1}} \otimes
         \left( \begin{array}{c}
		x_{n_a} \\
		x_{n_b} 
		\end{array} \right) \\ \hline
	x_{1_b}~\ket{x_2, \ldots, x_{n-1}} \otimes
         \left( \begin{array}{c}
		x_{n_a} \\
		x_{n_b} 
		\end{array} \right)
 \end{array}
  \right)=\\
 & = & \left(
  \begin{array}{c}
        {P_0^{(n-1)}} \cdot 
         x_{1_a}~\ket{x_2, \ldots, x_{n-1}} \otimes
         \left( \begin{array}{c}
		x_{n_a} \\
		x_{n_b} 
		\end{array} 
	\right)	
	~+~
	{L_1^{(n-1)}} \cdot
	x_{1_b}~\ket{x_2, \ldots, x_{n-1}} \otimes
         \left( \begin{array}{c}
		x_{n_a} \\
		x_{n_b} 
		\end{array} \right)	
		\\ \hline
	{L_0^{(n-1)}} \cdot 
         x_{1_a}~\ket{x_2, \ldots, x_{n-1}} \otimes
         \left( \begin{array}{c}
		x_{n_a} \\
		x_{n_b} 
		\end{array} 
	\right)	
	~+~
	{P_1^{(n-1)}} \cdot
	x_{1_b}~\ket{x_2, \ldots, x_{n-1}} \otimes
         \left( \begin{array}{c}
		x_{n_a} \\
		x_{n_b} 
		\end{array} \right)
 \end{array}
  \right)=\\
 & = & \left(
  \begin{array}{c}
         {I^{{(n-2)}}} \cdot 
         x_{1_a}~\ket{x_2, \ldots, x_{n-1}} \otimes
         {P_0} \cdot
         \left( \begin{array}{c}
		x_{n_a} \\
		x_{n_b} 
		\end{array} 
	\right)	
	~+~
	{I^{{(n-2)}}} \cdot
	x_{1_b}~\ket{x_2, \ldots, x_{n-1}} \otimes
	{L_1}\cdot
         \left( \begin{array}{c}
		x_{n_a} \\
		x_{n_b} 
		\end{array} \right)	
		\\ \hline
	{I^{{(n-2)}}} \cdot 
         x_{1_a}~\ket{x_2, \ldots, x_{n-1}} \otimes
         {L_0}\cdot
         \left( \begin{array}{c}
		x_{n_a} \\
		x_{n_b} 
		\end{array} 
	\right)	
	~+~
	{I^{{(n-2)}}} \cdot
	x_{1_b}~\ket{x_2, \ldots, x_{n-1}} \otimes
	{P_1} \cdot
         \left( \begin{array}{c}
		x_{n_a} \\
		x_{n_b} 
		\end{array} \right)
 \end{array}
  \right)=\\
 & = &  \left( 
          \begin{array}{c}
           \left( 
          \begin{array}{c}
            x_{1_a} x_{n_a} \ket{x_2, \ldots, x_{n-1}} \\
            x_{1_b} x_{n_a} \ket{x_2, \ldots, x_{n-1}} 
            \end{array}
         \right) \\          \hline
          \left( 
          \begin{array}{c}
             x_{1_a} x_{n_b} \ket{x_2, \ldots, x_{n-1}} \\
            x_{1_b} x_{n_b} \ket{x_2, \ldots, x_{n-1}}
            \end{array}
         \right) 
          \end{array}
         \right) =    \\      
 & = & \left( 
	  \begin{array}{c}
		x_{n_a} \\
		x_{n_b} 
	   \end{array} 
     \right) 
     \otimes 
     \ket{x_2, \ldots, x_{n-1}} 
     \otimes
     \left( 
	  \begin{array}{c}
		x_{1_a} \\
		x_{1_b} 
	   \end{array} 
     \right)= \\
 & = & \ket{x_n} \otimes \ket{x_2, \ldots, x_{n-1}} \otimes \ket{x_1}.
 \end{eqnarray*}

To check the unitarity of $SWAP_{(n;1,n)}$, first
it is easy to see that $(P_0^{(n)})^{\dagger}=P_0^{(n)}$ and $(P_1^{(n)})^{\dagger}=P_1^{(n)}.$ Further, $(L_0^{(n)})^{\dagger}=(I^{(n-1)}\otimes L_0)^{\dagger}=I^{(n-1)}\otimes L_0^{\dagger}=I^{(n-1)}\otimes L_1=L_1^{(n)}.$ 

Similarly, $(L_1^{(n)})^{\dagger}=L_0^{(n)},$ hence $(SWAP_{(n;1,,n)})^{\dagger}= SWAP_{(n;1,,n)}; SWAP_{(n;1,,n)}\cdot (SWAP_{(n;1,,n)})^{\dagger}=(SWAP_{(n;1,,n)})^{\dagger}\cdot SWAP_{(n;1,,n)}=I^{(n)}.$

\bigskip

Now it is straightforward to consider a synthetic mathematical representation of $SWAP_{(k;m,m+n)}$ (depicted in Figure 3) that performs a swapping between the $m$-th and the $(m+n)$-th qubits of a $k$-dimensional input state. We simply consider to keep the first $(m-1)$ and the last $(k-m-n-1)$ qubits unchanged and apply the $SWAP_{(n;1,n)}$ gate to the state $\ket{x_m,\cdots,x_{m+n}}.$ Formally, we end up with the following theorem. 

\begin{theorem}\label{gswap}
\nl
Let us consider $\ket{x_1,\cdots, x_m, \cdots x_{m+n}, \cdots, x_k}$ such that $\ket {x_i}\in \mathbb C^2$ $\forall i\in\{1,\cdots,k\}.$ Then, $$SWAP_{(k;m,m+n)}=I^{(m-1)}\otimes SWAP_{(n+1;1,n+1)}\otimes I^{(k-m-n)}.$$
\end{theorem}

Proof:

\begin{eqnarray*}
&& SWAP_{(k;m,m+n)}\ket{x_1,\cdots, x_m, \cdots, x_{m+n}, \cdots x_k}= \\
&=& (I^{(m-1)}\ket{x_1,\cdots, x_{m-1}})\otimes (SWAP_{(n+1;1,n+1)}\ket{x_m,\cdots, x_{m+n}})\\
&\otimes& (I^{(k-m-n)}\ket{x_{m+n+1},\cdots, x_k})=\\
&=& \ket{x_1,\cdots, x_{m+n}, \cdots, x_{m}, \cdots, x_k}.
\end{eqnarray*}

\section{The block-matrix representation of a binary gate}\label{4}
Let us consider an arbitrary binary gate $U^{(2)}$. If the dimension of the input state $\ket{x_1,\cdots,x_k}$ is $k$ and the gate $U^{(2)}$ is applied to two consecutive qubits $\ket {x_n}$ and $\ket {x_{n+1}}$, then the block-matrix expression of $U^{(2)}$ applied to $\ket{x_1,\cdots,x_k}$ is simply $I^{n-1}\otimes U^{(2)}\otimes I^{k-n-1}.$ 
In the previous Section we have provided a block-matrix representation of the SWAP gate applied to two arbitrary qubits. Similarly, in this Section we consider the general case where an arbitrary binary gate $U^{(2)}$ is applied to two arbitrary qubits (generally non-adjacent) of a $k$-dimensional input state, as depicted in Figure 1 for the particular case of $SWAP_{(k;m,m+n)}$. In particular, we name a binary gate $U^{(2)}$ that is applied to the $m$-th and the $(m+n)$-th qubits of a $k$-dimensional input state as $U_{(k;m,m+n)}$. 

\begin{figure}[htbp]\label{Binary}
\centering
\includegraphics[scale=.18]{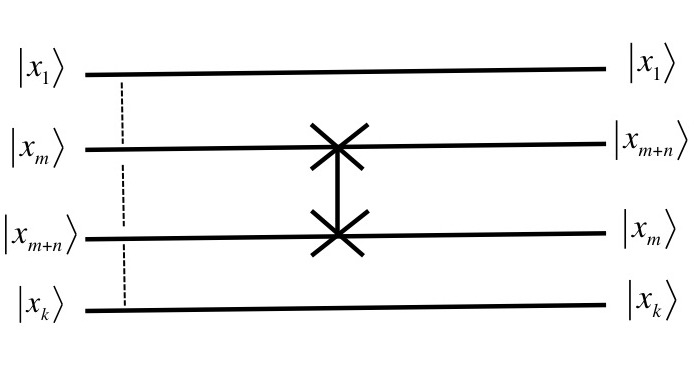}
\caption{Circuit representation of the $SWAP_{(k;m,m+n)}$ gate.}
\end{figure}

The usual strategy \cite{BK} to apply the gate $U^{(2)}$ to the $m$-th and the $(m+n)$-th qubits, consists into performing  multiple $SWAP$ gates in order to arrange the two qubits $\ket{x_m}$ and $\ket{x_{m+n}}$ in two adjacent positions. Then, the gate $U^{(2)}$ is applied; after it is necessary to apply again  multiple $SWAP$ gates to retrieve the circuit at the initial order of all the qubits. The suitable application of multiple $SWAP$ gates is determined by simple permutations of qubits; in particular, to apply a binary gate $U^{(2)}$ between the $\ket{x_m}$-th and the $\ket{x_{m+n}}$-th qubits, it is necessary to perform $2n$ $SWAP$ of consecutive qubits.

As an example, let us consider to apply the $C-Not$ gate to the second and the fifth qubits over a circuit of six qubits. Following tha standard strategy:
\begin{itemize} 
\item we apply the $SWAP_{(6;4,5)}$ between the $4$-th and the $5$-th qubit;
\item we apply the $SWAP_{(6;3,4)}$ between the $3$-th and the $4$-th qubit. In this way, the qubit that was originally in the $5$-th position, comes in the $3$-th position (i.e. adjacent the $2$-nd qubit);
\item now we apply the operator $I\otimes (C-Not)\otimes I^{(2)}$;
\item finally we apply $SWAP_{(6;3,4)}^{-1}\cdot SWAP_{(6;4,5)}^{-1}$ to retrive the original configuration.
\end{itemize}

Hence, the mathematical form of this operation is given by the following composition:
$$SWAP_{(6;4,5)}\cdot SWAP_{(6;3,4)}\cdot I\otimes (C-Not)\otimes I^{(2)}\cdot SWAP_{(6;3,4)}^{-1}\cdot SWAP_{(6;4,5)}^{-1}.$$
It is easy to realize how this procedure could be generalized also for arbitrary $n$-ary gates but producing very complex composition of many $SWAP$ gates (as detailed in Section \ref{MS}). On this basis, from a mathematical point of view a more synthetic representation may be desirable.

The result given by Theorem (\ref{gswap}) allows to provide a block-matrix representation of an arbitrary binary gate $U_{(k;m,m+n)}$ applied to the $m$-th and $(m+n)$-th qubits of a $k$-dimensional input state.

Let $U^{(2)}$ a binary unitary operator given by the following block-matrix representation $
U^{(2)} =  \left[ \begin{array}{c|c}
					U_{11} & U_{12} \\ \hline          
					  U_{21}&  U_{22} \\ 
						\end{array}\right],
$ where $U_{ij}$ are $2-$ dimensional square matrices given by $U_{11} =  \left[ \begin{array}{cc}
					u_{11} & u_{12} \\           
					  u_{21}&  u_{22} \\ 
						\end{array}\right],$ $ U_{12} =  \left[ \begin{array}{cc}
					u_{13} & u_{14} \\           
					  u_{23}&  u_{24} \\ 
						\end{array}\right]$ , $ U_{21} =  \left[ \begin{array}{cc}
					u_{31} & u_{32} \\           
					  u_{41}&  u_{42} \\ 
						\end{array}\right]$ and $ U_{22} =  \left[ \begin{array}{cc}
					u_{23} & u_{24} \\           
					  u_{43}&  u_{44} \\ 
						\end{array}\right].$ 

\begin{theorem}\label{U}

The block-matrix representation of $U_{(k;m,m+n)}$ is given by: $$U_{(k;m,m+n)}=I^{(m-1)}\otimes \left[ \begin{array}{c|c}
					U_{11}^{(n)} & U_{12}^{(n)} \\  \hline         
					  U_{21}^{(n)}&  U_{22}^{(n)} \\ 
					\end{array}\right]\otimes I^{(k-m-n)},$$
where $U_{ij}^{(n)}=I^{(n-1)}\otimes U_{ij}.$
\end{theorem}

Proof:

The strategy is based to
\begin{enumerate} 
\item apply the $SWAP_{(k;m+1,m+n)}$ in order to ``place" the $(m+n)$-th qubit in the $(m+1)$-th position;
\item apply the  binary gate $U^{(2)}$ to the $m$-th and the $m+1$-th qubits;
\item apply again the $SWAP_{(k;m+1,m+n)}$ in order to recover the initial disposition of the qubits.
\end{enumerate}

Fomally,

\begin{eqnarray*}
& U_{(k;m,m+n)}=\\
&= SWAP_{(k;m+1,m+n)}\cdot (I^{(m-1)}\otimes U^{(2)} \otimes I^{(k-m-1)})\cdot SWAP_{(k;m+1,m+n)}=\\
&=(I^{(m)}\otimes SWAP_{(n+1;1,n+1)}\otimes I^{(k-m-n-1)})\cdot \\
& (I^{(m-1)}\otimes U^{(2)} \otimes I^{(k-m-1)})\cdot \\
& (I^{(m)}\otimes SWAP_{(n+1;1,n+1)}\otimes I^{(k-m-n-1)})=\\
&=(I^{(m-1)}\otimes(I\otimes SWAP_{(n+1;1,n+1)})\otimes I^{(k-m-n-1)})\cdot \\
& (I^{(m-1)}\otimes U^{(2)} \otimes I^{(k-m-1)})\cdot \\
& (I^{(m-1)}\otimes(I\otimes SWAP_{(n+1;1,n+1)})\otimes I^{(k-m-n-1)})=\\
& I^{(m-1)}\otimes ((I\otimes SWAP_{(n+1;1,n+1)})\cdot(U^{(2)}\otimes I^{(n)})\cdot (I\otimes SWAP_{(n+1;1,n+1)}))\otimes I^{(k-m-n-1)}.
\end{eqnarray*}

Let us remark that $$\left[ \begin{array}{c|c}
					U_{11}\otimes I^{(n)} & U_{12}\otimes I^{(n)} \\  \hline         
	U_{21}\otimes I^{(n)}				  &  U_{22}\otimes I^{(n)}  \\ 
					\end{array}\right]=\left[ \begin{array}{c|c}
					U_{11} & U_{12} \\  \hline         
	U_{21}				  &  U_{22}  \\ 
					\end{array}\right]\otimes I^{(n)}$$ but $$\left[ \begin{array}{c|c}
				I^{(n)}\otimes	U_{11} & I^{(n)}\otimes U_{12} \\  \hline         
	I^{(n)}\otimes U_{21}	  &  I^{(n)}\otimes U_{22} \\ 
					\end{array}\right]\neq I^{(n)}\otimes\left[ \begin{array}{c|c}
					U_{11} & U_{12} \\  \hline         
	U_{21}				  &  U_{22}  \\ 
					\end{array}\right].$$

Hence, 
\begin{eqnarray*}
&  (I\otimes SWAP_{(n+1;1,n+1)})\cdot(U\otimes I^{(n)})\cdot (I\otimes SWAP_{(n+1;1,n+1)})=\\
&=  \left[ \begin{array}{c|c}
					SWAP_{(n+1;1,n+1)} & 0 \\  \hline         
					0  &  SWAP_{(n+1;1,n+1)} \\ 
					\end{array}\right]\cdot \left[ \begin{array}{c|c}
					U_{11}\otimes I^{(n)} & U_{12}\otimes I^{(n)} \\  \hline         
	U_{21}\otimes I^{(n)}				  &  U_{22}\otimes I^{(n)}  \\ 
					\end{array}\right] \cdot\\ &\left[ \begin{array}{c|c}
					SWAP_{(n+1;1,n+1)} & 0 \\  \hline         
					 0 &  SWAP_{(n+1;1,n+1)} \\ 
					\end{array}\right]=\left[ \begin{array}{c|c}
					\tilde{U}_{11} & \tilde{U}_{12}  \\  \hline         
	\tilde{U}_{21} 				  &  \tilde{U}_{22} \\ 
					\end{array}\right]
\end{eqnarray*} 

where, for the sake of the brevity, we place: $\tilde{U}_{ij}=SWAP_{(n+1;1,n+1)}\cdot (U_{ij}\otimes I^{(n)})\cdot SWAP_{(n+1;1,n+1)}.$
Let us notice that for any pair of unitary operators $S^{(m)}$ (of dimension $m$) and $T^{(n)}$ (of dimension $n$), trivially follows that:  $$SWAP_{(k;m,m+n)}\cdot(S^{(m)}\otimes T^{(n)})\cdot SWAP_{(k;m,m+n)}=T^{(n)}\otimes S^{(m)}.$$ Then, we have our claim.

Further, given the unitarity of $U^{(2)}$ and reminding that the tensor product preserves the unitarity \cite{F}, the unitarity of $U_{(k;m,m+n)}$ is given by construction.

\bigskip

\section{An Example: the $\sqrt{SWAP}_{(k;m,m+n)}$}

As an example, in this subsection we show an application of the Theorem \ref{U} in order to provide a block-matrix representation of the Square Root of $SWAP$ gate $\sqrt{SWAP}_{(k;m,m+n)}$.

The standard $\sqrt{SWAP}$ gate is a binary unitary operator whose matrix expression is given by:
$$\sqrt{SWAP}=\left[ \begin{array}{cccc}
					 1 & 0  & 0 & 0 \\    
					 0 & \frac{1+i}{2}  & \frac{1-i}{2} & 0 \\ 
             0 & \frac{1-i}{2}  & \frac{1+i}{2} & 0 \\ 
             0 & 0  & 0 & 1 \\ 
						\end{array}\right]$$
that represents a gate such that, if it is applied two times to a pair of qubits, it swaps these qubits. Interestingly enough, the $\sqrt{SWAP}$ is well known as a kind of ``entangling" gate, i.e. a gate that applied to a vector of the computational basis, gives an entangled state as output \cite{GLSP}.
Let us consider to apply $\sqrt{SWAP}$ to the $m$-th and the $(m+n)$-th qubits of a $k$-dimensional input state. 

At this aim and in accord with the Theorem \ref{U}, let us introduce the following unary matrices:

$SS_{11} =  \left[ \begin{array}{cc}
					1 & 0 \\           
					    0 & \frac{1+i}{2}\\ 
						\end{array}\right]$ , $ SS_{12} =  \left[ \begin{array}{cc}
					0 & 0 \\           
					  \frac{1-i}{2}&  0 \\ 
						\end{array}\right]$ , $ SS_{21} =  \left[ \begin{array}{cc}
					0 & \frac{1-i}{2} \\           
					  0 & 0 \\ 
						\end{array}\right]$ and\\ $ SS_{22} =  \left[ \begin{array}{cc}
					\frac{1+i}{2} & 0 \\           
					  0 &  1 \\ 
						\end{array}\right]$ such that $
\sqrt{SWAP} =  \left[ \begin{array}{c|c}
					SS_{11} & SS_{12} \\ \hline          
					  SS_{21}&  SS_{22} \\ 
						\end{array}\right].
$ By appealing to the Theorem \ref{U}, $$\sqrt{SWAP}_{(k;m,m+n)}=I^{(m-1)}\otimes \left[ \begin{array}{c|c}
					SS_{11}^{(n)} & SS_{12}^{(n)} \\  \hline         
					  SS_{21}^{(n)}&  SS_{22}^{(n)} \\ 
					\end{array}\right]\otimes I^{(k-m-n)},$$ where $SS_{ij}^{(n)}=I^{(n-1)}\otimes SS_{ij}.$
Let us verify that: 
\begin{enumerate}
\item $\sqrt{SWAP}_{(k;m,m+n)}\cdot \sqrt{SWAP}_{(k;m,m+n)}={SWAP}_{(k;m,m+n)};$
\item $\sqrt{SWAP}_{(k;m,m+n)}$ is unitary.
\end{enumerate} 

\bigskip

\begin{enumerate}
\item 
\begin{eqnarray*}
&\sqrt{SWAP}_{(k;m,m+n)}\cdot \sqrt{SWAP}_{(k;m,m+n)}=\\
&=I^{(m-1)}\otimes \left[ \begin{array}{c|c}
					SS_{11}^{(n)} & SS_{12}^{(n)} \\  \hline         
					  SS_{21}^{(n)}&  SS_{22}^{(n)} \\ 
\end{array}\right]\cdot \left[ \begin{array}{c|c}
					SS_{11}^{(n)} & SS_{12}^{(n)} \\  \hline         
					  SS_{21}^{(n)}&  SS_{22}^{(n)} \\ 
\end{array}\right] \otimes I^{(k-m-n)}=\\
&=I^{(m-1)}\otimes \left[ \begin{array}{c|c}
					\alpha_{11} & \alpha_{12} \\  \hline         
					  \alpha_{21}&  \alpha_{22} \\ 
\end{array}\right] \otimes I^{(k-m-n)},
\end{eqnarray*}
where 

$\alpha_{11}=SS_{11}^{(n)}\cdot SS_{11}^{(n)}+ SS_{12}^{(n)}\cdot SS_{21}^{(n)}=$ $I^{(n-1)}\otimes (SS_{11}\cdot SS_{11}+SS_{12}\cdot SS_{21})=I^{(n-1)}\otimes P_0=P_0^{(n)}.$

Analogously,

$\alpha_{12}=SS_{11}^{(n)}\cdot SS_{12}^{(n)}+ SS_{12}^{(n)}\cdot SS_{22}^{(n)}=$ $I^{(n-1)}\otimes (SS_{11}\cdot SS_{12}+SS_{12}\cdot SS_{22})=I^{(n-1)}\otimes L_1=L_1^{(n)}.$

$\alpha_{21}=SS_{21}^{(n)}\cdot SS_{11}^{(n)}+ SS_{22}^{(n)}\cdot SS_{21}^{(n)}=$ $I^{(n-1)}\otimes (SS_{21}\cdot SS_{11}+SS_{22}\cdot SS_{21})=I^{(n-1)}\otimes L_0=L_0^{(n)}.$

$\alpha_{11}=SS_{21}^{(n)}\cdot SS_{12}^{(n)}+ SS_{22}^{(n)}\cdot SS_{22}^{(n)}=$ $I^{(n-1)}\otimes (SS_{21}\cdot SS_{12}+SS_{22}\cdot SS_{22})=I^{(n-1)}\otimes P_1=P_1^{(n)}.$

\item To prove the unitarity of $\sqrt{SWAP}_{(k;m,m+n)}$ we need to prove that $$\left[ \begin{array}{c|c}
					SS_{11}^{(n)} & SS_{12}^{(n)} \\  \hline         
					  SS_{21}^{(n)}&  SS_{22}^{(n)} \\ 
\end{array}\right]^{\dagger}\cdot\left[ \begin{array}{c|c}
					SS_{11}^{(n)} & SS_{12}^{(n)} \\  \hline         
					  SS_{21}^{(n)}&  SS_{22}^{(n)} \\ 
\end{array}\right]=\left[ \begin{array}{c|c}
					SS_{11}^{(n)} & SS_{12}^{(n)} \\  \hline         
					  SS_{21}^{(n)}&  SS_{22}^{(n)} \\ 
\end{array}\right]\cdot\left[ \begin{array}{c|c}
					SS_{11}^{(n)} & SS_{12}^{(n)} \\  \hline         
					  SS_{21}^{(n)}&  SS_{22}^{(n)} \\ 
\end{array}\right]^{\dagger}=I^{(n+1)}.$$
Let us recall that the transpose of an arbitrary block matrix $\left[ \begin{array}{c|c}
					A_{11} & A_{12} \\ \hline     
					 A_{21} & A_{22} \\ 
						\end{array}\right]$ is given by $\left[ \begin{array}{c|c}
					A_{11} & A_{21}^t \\ \hline     
					 A_{12}^t & A_{22} \\ 
						\end{array}\right]$. Further, it is easy to see that $SS_{12}^t=SS_{21}.$ Hence, we have:

\begin{eqnarray*}
&\left[ \begin{array}{c|c}
					SS_{11}^{(n)} & SS_{12}^{(n)} \\  \hline         
					  SS_{21}^{(n)}&  SS_{22}^{(n)} \\ 
\end{array}\right]^{\dagger}\cdot\left[ \begin{array}{c|c}
					SS_{11}^{(n)} & SS_{12}^{(n)} \\  \hline         
					  SS_{21}^{(n)}&  SS_{22}^{(n)} \\ 
\end{array}\right]=\\
&=\left[ \begin{array}{c|c}
					(SS_{11}^{(n)})^{\dagger} & (SS_{12}^{(n)})^{\dagger} \\  \hline         
					  (SS_{21}^{(n)})^{\dagger}&  (SS_{22}^{(n)})^{\dagger} \\ 
\end{array}\right]\cdot\left[ \begin{array}{c|c}
					SS_{11}^{(n)} & SS_{12}^{(n)} \\  \hline         
					  SS_{21}^{(n)}&  SS_{22}^{(n)} \\ 
\end{array}\right]=\left[ \begin{array}{c|c}
					A_{11} & A_{12} \\  \hline         
					  A_{21} &  A_{12} \\ 
\end{array}\right]
\end{eqnarray*}
where:

$A_{11}=(S_{11}^{(n)})^{\dagger}\cdot S_{11}^{(n)}+(S_{12}^{(n)})^{\dagger}\cdot S_{21}^{(n)}=I^{(n)}\cdot((S_{11})^{\dagger}\cdot S_{11}+(S_{12})^{\dagger}\cdot S_{21})=I^{(n)}\cdot I^{(n)}=I^{(n)};$

$A_{12}=(S_{11}^{(n)})^{\dagger}\cdot S_{12}^{(n)}+(S_{21}^{(n)})^{\dagger}\cdot S_{22}^{(n)}=I^{(n)}\cdot((S_{11})^{\dagger}\cdot S_{12}+(S_{12})^{\dagger}\cdot S_{22})=I^{(n)}\cdot \bf{0}=0$, where $\bf{0}$ is the null matrix. Analogously, 

$A_{21}=(S_{21}^{(n)})^{\dagger}\cdot S_{11}^{(n)}+(S_{22}^{(n)})^{\dagger}\cdot S_{21}^{(n)}=I^{(n)}\cdot((S_{21})^{\dagger}\cdot S_{11}+(S_{22})^*\cdot S_{21})=I^{(n)}\cdot \bf{0}=0.$

$A_{22}=(S_{21}^{(n)})^{\dagger}\cdot S_{12}^{(n)}+(S_{22}^{(n)})^{\dagger}\cdot S_{22}^{(n)}=I^{(n)}\cdot((S_{21})^{\dagger}\cdot S_{12}+(S_{22})^{\dagger}\cdot S_{22})=I^{(n)}\cdot I^{(n)}=I^{(n)}.$

In a very similar way, it can be also checked that $$\left[ \begin{array}{c|c}
					SS_{11}^{(n)} & SS_{12}^{(n)} \\  \hline         
					  SS_{21}^{(n)}&  SS_{22}^{(n)} \\ 
\end{array}\right]\cdot\left[ \begin{array}{c|c}
					SS_{11}^{(n)} & SS_{12}^{(n)} \\  \hline         
					  SS_{21}^{(n)}&  SS_{22}^{(n)} \\ 
\end{array}\right]^{\dagger}=I^{(n+1)}.$$
\end{enumerate} 

Let us notice that the application of the $\sqrt{SWAP}_{(k;m,m+n)}$ produces an entangled state between the $m$-th and the $(m+n)$-th qubits. Without lost of generality, it is possible to apply another arbitrary binary operator $U_{(k;m,m+n)}$ to the entangled state generated by the previous application of the  $\sqrt{SWAP}_{(k;m,m+n)}$ and we end up with the following representation:

$$(I^{(m-1)}\otimes\sqrt{SWAP}_{(k;m,m+n)}\otimes I^{(k-m-n)})\cdot (I^{(m-1)}\otimes U_{(k;m,m+n)}\otimes I^{(k-m-n)})=$$
$$=I^{(m-1)}\otimes(\sqrt{SWAP}_{(k;m,m+n)}\cdot U_{(k;m,m+n)})\otimes I^{(k-m-n)}).$$

\section{Multiple Swapping}\label{MS}
In Section 4 we have showed how a block-matrix representation of the $SWAP$ gate between two arbitrary (generally non-adjacent) qubits belonging to a $k$-dimensional input state, allows to provide a block-matrix representation of an arbitrary binary unitary operator.
Let us notice that, without any loss of generality, the strategy adopted in Theorem \ref{U} can be easily generalized to an arbitrary $n$-ary unitary operator $U^{(n)}$; the expedient simply consists in dividing the matrix $U^{(n)}$ in four block matrices, each one of dimension $2^{n-1}$ (instead of $2$, as in the binary case). But, in order to apply an $n$-ary unitary operator to $n$ arbitrary (generally non adjacent) qubits, we first need to perform multiple swap among the qubits of the input state. 

In other words, let us consider to apply an $n$-ary quantum gate $U^{(n)}$ to the $\alpha_1$-th, $\alpha_2$-th,$\cdots \alpha_n$-th qubits of a $k$-dimensional input state. The $n$ qubits which the operator $U^{(n)}$ is applied to, may, in general, be not adjacent. Similarly to the binary case mentioned above, the standard strategy is based to consider all the $\alpha_2-(\alpha_1+1)$ permutations to swap the $\alpha_2$-th qubit to the $(\alpha_1+1)$-th entry, all the $\alpha_3-(\alpha_1+2)$ permutations to swap the $(\alpha_3)$-th qubit to the $(\alpha_1+2)$-th entry and so on. Following this procedure, it is necessary to consider $2\sum_{i=2}^n(\alpha_i-(\alpha_{1}+i-1))$   applications of binary SWAP gates to adjacent qubits in order to apply $U^{(n)}$ to the required qubits and to restore the circuit at the initial configuration.
However, Theorem \ref{U} allows to reduce this procedure providing a more simple expression of multiple $SWAP$ gate.
 
As an example, let us suppose to apply two different swap to a $k$-dimensional input state. Formally,

\begin{eqnarray*}
&SWAP_{(k;m,m+n)}\cdot SWAP_{(k;m',m'+n')}=\\
&=(I^{(m-1)}\otimes SWAP_{(n+1;1,n+1)}\otimes I^{(k-m-n)})\cdot \\
&\cdot(I^{(m'-1)}\otimes SWAP_{(n'+1;1,n'+1)}\otimes I^{(k-m'-n')})=
\end{eqnarray*}
without any loss of generality, we can assume that $m\geq m'$ and $n\geq n'$, hence:

\begin{eqnarray*}
&=(I^{(m'-1)}\otimes (I^{(m-m')}\otimes SWAP_{(n+1;1,n+1)})\otimes I^{(k-m-n)})\cdot \\
&\cdot(I^{(m'-1)}\otimes (SWAP_{(n'+1;1,n'+1)}\otimes I^{((m-m')+(n-n'))})\otimes I^{(k-m-n)})=\\
&=I^{(m'-1)}\otimes((I^{(m-m')}\otimes SWAP_{(n+1;1,n+1)})\cdot \\
&\cdot(SWAP_{(n'+1;1,n'+1)}\otimes I^{((m-m')+(n-n'))}))\otimes I^{(k-m-n)},
\end{eqnarray*}

where $SWAP_{(n+1;1,n+1)}$ and $SWAP_{(n'+1;1,n'+1)}$ follow the block-matrix representation given by Lemma \ref{swap}. This procedure allows to simply obtain the output of the application of a ternary gate to $3$ arbitrary qubits of a $k$-dimensional input state without performing a composition of $2\sum_{i=2}^3(\alpha_i-(\alpha_1+i-1))=2(\alpha_3+\alpha_2-2\alpha_1-3)$ binary $SWAP$ gates.

By following the same procedure, it is possible to determinate a synthetic form of arbitrary multiple $SWAP$ gate. This naturally allows to provide a very synthetic representation of an arbitrary $n$-ary quantum gate and, in principle, also of arbitrary sequences of quantum gates. By this perspective, a synthetic matrix representation of an arbitrary quantum circuit is easy to achieve by involving compositions of multiple $SWAP$ gates and arbitrary dimensional unitary operators. On the other hand, this representation avoid to incur into the annoying request to consider the composition of several binary $SWAP$ gates coming from the arguments discussed above.
 These examples aim to suggest that possible advantages of this simplification could tourn out to be very effective in case of simulating quantum circuits by using a classical programming language \cite{Gerdt1,Gerdt2,Gerdt3}, mostly in case of many-qubits (or even many-qudits) quantum circuits. Indeed, this kind of simplification could be extremely useful in designing a classical software package devoted to simulate complex quantum circuits.    

\section{A block-matrix  representation of Quantum Circuits with qudits} 
As it is well known, the qubit is the basic concept of the quantum information theory. A natural many-valued generalization of the qubit is represented by the qudit, that is a unit vector in the Hilbert space $\mathbb C^d$. In principle, it is possible to think a quantum circuit where the input state is given by a register of $k$ qudits instead of $k$ qubits, as has been recently considered by several authors \cite{TNWM,ZLW}. In this Section, we provide a generalization of the Theorem \ref{U} in the framework of quantum circuits with qudits.

Let us indicate by $\mathbb B^{(d)}=\{\ket{b_0},\ket{b_1},\cdots,\ket{b_{d-1}}\}$ the computational basis of $\mathbb C^d$, where the \emph{qudit} $\ket{b_i}=\ket{\frac{i}{d-1}}=\left[\begin{array}{c}
					0  \\  
					  0   \\ 
\vdots \\
1 \\
\vdots \\
0
\end{array}\right]$ is a vector with 1 in the $(i+1)$-th entry and $0$ in all the other entries and let us also define the quantity $_{[d]}P_{(j,k)}=\ket{b_j}\langle{b_k}|.$ First, let us provide a block-matrix representation of the binary $SWAP$ gate in the framework of quantum circuits with qudits.
\begin{theorem}
Let consider two qudits $\ket{x},\ket y \in \mathbb C^d$ and let $\mathbb B^{d}$ the computational basis on $\mathbb C^d.$ Then, the SWAP gate $_{[d]}SWAP$ between two (adjacent) qudits is given by:

$$_{[d]}SWAP= \left[ \begin{array}{c|c|c|c}
					 _{[d]}P_{(0,0)} & _{[d]}P_{(1,0)}  & \cdots & _{[d]}P_{(d-1,0)} \\ \hline   
					 _{[d]}P_{(0,1)} & _{[d]}P_{(1,1)}  & \cdots & _{[d]}P_{(d-1,1)} \\ \hline
             \vdots &   & \ddots &  \\  \hline
             _{[d]}P_{(0,d-1)} & _{[d]}P_{(1,d-1)}  & \cdots & _{[d]}P_{(d-1,d-1)} \\ 
						\end{array}\right].$$
\end{theorem}

Proof:

We need to prove that
 {$_{[d]}SWAP(\ket x \otimes \ket y)=\ket y \otimes \ket x$}
 and to verify the unitarity of $_{[d]}SWAP$.

\begin{enumerate}
\item{
Let $\ket x=\left[ \begin{array}{c}
					x_0  \\    
					 x_1 \\ 
             \vdots  \\ 
              x_{d-1}\\ 
						\end{array}\right]$ and $\ket y=\left[ \begin{array}{c}
					y_0  \\    
					 y_1 \\ 
             \vdots  \\ 
              y_{d-1}\\ 
						\end{array}\right]$ two arbitray qudits.
\begin{eqnarray*}
&_{[d]}SWAP(\ket x \otimes \ket y)=\left[ \begin{array}{c}
					 {x_{0}}_{[d]}P_{(0,0)}\ket y +{x_{1}}_{[d]}P_{(1,0)}\ket y +\cdots+{x_{d-1}}_{[d]}P_{(d-1,0)}\ket y\\    \hline
					 {x_{0}}_{[d]}P_{(0,1)}\ket y +{x_{1}}_{[d]}P_{(1,1)}\ket y +\cdots+{x_{d-1}}_{[d]}P_{(d-1,1)}\ket y \\ \hline
             \vdots  \\  \hline
              {x_{0}}_{[d]}P_{(0,d-1)}\ket y +{x_{1}}_{[d]}P_{(1,d-1)}\ket y +\cdots+{x_{d-1}}_{[d]}P_{(d-1,d-1)}\ket y\\ 
						\end{array}\right].
\end{eqnarray*}
By construction, $_{[d]}P_{(j,k)}$ has $1$ in the $(j+1,k+1)$-th entry and $0$ everywhere else. Hence,

\begin{eqnarray*}
& _{[d]}SWAP(\ket x \otimes \ket y)=\left[ \begin{array}{c}
					 \left[ \begin{array}{cccc}
					 x_0 & 0 & \cdots & 0\\    
             x_1 & 0 & \cdots &   \\ 
              \vdots &  & \ddots & 0\\
x_{d-1} & 0 & \cdots & 0   
						\end{array}\right]\ket y\\    
             \left[ \begin{array}{cccc}
					 0 & x_0 & \cdots & 0\\    
             0 & x_1 & \cdots &   \\ 
              \vdots &  & \ddots & 0\\
0 & x_{d-1} & \cdots & 0   
						\end{array}\right]\ket y  \\ 
              \vdots\\
\left[ \begin{array}{cccc}
					 0 & 0 & \cdots & x_0\\    
             0 & 0 & \cdots &  x_1 \\ 
              \vdots &  & \ddots & 0\\
0 & 0 & \cdots & x_{d-1}   
						\end{array}\right]\ket y
						\end{array}\right]= \left[ \begin{array}{c}
					 \left[ \begin{array}{c}
					 x_0\cdot y_0 \\    
             x_1 \cdot y_0    \\ 
              \vdots \\
x_{d-1} \cdot y_0 
						\end{array}\right]\\    
             \left[ \begin{array}{c}
					 x_0 \cdot y_1\\    
             x_1 \cdot y_1   \\ 
              \vdots \\
x_{d-1}  \cdot y_1
						\end{array}\right]\\ 
              \vdots\\
\left[ \begin{array}{c}
					 x_0 \cdot y_{d-1}\\    
             x_1  \cdot y_{d-1}  \\ 
              \vdots \\
x_{d-1} \cdot y_{d-1} 
						\end{array}\right]
						\end{array}\right]=\ket y \otimes \ket x.
\end{eqnarray*}
\item $_{[d]}SWAP^{\dagger}=\left[ \begin{array}{c|c|c|c}
					 (P_{(0,0)})^{\dagger} & (P_{(0,1)})^{\dagger} & \cdots & (P_{(0,d-1)})^{\dagger}\\   \hline 
(P_{(1,0)})^{\dagger} & (P_{(1,1)})^{\dagger} & \cdots & (P_{(1,d-1)})^{\dagger}\\ \hline
              \vdots \\ \hline
(P_{(d-1,0)})^{\dagger} & (P_{(d-1,1)})^{\dagger} & \cdots & (P_{(d-1,d-1)})^{\dagger}
						\end{array}\right],$

but the unitarity easily follows by noticing that $(P_{(i,j)})^{\dagger}=(P_{(i,j)})$     $\forall i,j\in\{0,\cdots,d-1\}.$

}
\end{enumerate}

By replacing the reasoning given in Lemma \ref{swap} and in Theorem \ref{gswap}, it is possible to obtain the general form of the binary $SWAP$ gate for non-adjacent qudits.
\begin{lemma}
Let us consider an input state belonging to the $d^n$-dimensional Hilbert space  $\otimes^n\mathbb C^{d}$ (i.e. an input state given by $n$ qudits). The $SWAP$ gate between the first and the $n$-th qudits assumes the following block-matrix form:

$$_{[d]}SWAP_{(n;1,n+1)}= \left[ \begin{array}{c|c|c|c}
					 _{[d]}P_{(0,0)}^{(n-1)} & _{[d]}P_{(1,0)}^{(n-1)}  & \cdots & _{[d]}P_{(d-1,0)}^{(n-1)} \\   \hline 
					 _{[d]}P_{(0,1)}^{(n-1)} & _{[d]}P_{(1,1)}^{(n-1)}  & \cdots & _{[d]}P_{(d-1,1)}^{(n-1)} \\ \hline
             \vdots &   & \ddots &  \\  \hline
             _{[d]}P_{(0,d-1)}^{(n-1)} & _{[d]}P_{(1,d-1)}^{(n-1)}  & \cdots & _{[d]}P_{(d-1,d-1)}^{(n-1)} \\ 
						\end{array}\right],$$ where $_{[d]}P_{(i,j)}^{(n-1)}=I^{(n-2)}\otimes _{[d]}P_{(i,j)}$ and $I^{(n-2)}$ indicates the $d^{(n-2)}$- identity matrix.
\end{lemma}
\begin{theorem}
Let us consider an input state in $\otimes^k\mathbb C^d$ (i.e. an input state given by $k$ qudits). The gate able to $SWAP$ the $m$-th and the $m+n$-th qudits assumes the following form:
$$_{[d]}SWAP_{(k;m,m+n)}=I^{(m-1)}\otimes _{[d]}SWAP_{(n+1;1,n+1)}\otimes I^{(k-m-n)}.$$
\end{theorem}

This allows to naturally provide a generalization of the Theorem \ref{U} in the contest of quantum circuits with qudits.

Let $T$ be a binary operator for qudits and let us consider the block-matrix representation of $U$ as:
$$T=\left[ \begin{array}{c|c|c|c}
					 T_{11} & T_{12}  & \cdots & T_{1d} \\   \hline 
					 T_{21} & T_{22}  & \cdots & T_{2d} \\ \hline
             \vdots &   & \ddots &  \\  \hline
             T_{d1} & T_{d2}  & \cdots & T_{dd} \\ 
						\end{array}\right],$$ where $T_{ij}$ are $d$-dimensional square blocks of $T$.

\begin{theorem}\label{QT}
Let $T$ a binary operator for qudits and let us consider to apply $T$ to the $m$-th and the $(m+n)$-th qudits of a $k$-dimensinal input. The block-matrix representation of $T_{(k;m,m+n)}$ is given by: $$T_{(k;m,m+n)}=I^{(m-1)}\otimes \left[ \begin{array}{c|c|c|c}
					 T_{11}^{(n)} & T_{12}^{(n)}  & \cdots & T_{1d}^{(n)} \\    \hline
					 T_{21}^{(n)} & T_{22}^{(n)}  & \cdots & T_{2d}^{(n)} \\  \hline
             \vdots &   & \ddots &  \\ \hline 
             T_{d1}^{(n)} & T_{d2}^{(n)}  & \cdots & T_{dd}^{(n)} \\ 
					\end{array}\right]\otimes I^{(k-m-n)},$$

where $T_{ij}^{(n)}=I^{(n-1)}\otimes T_{ij}$ and $I^{(n-1)}$ is the $d^{(n-1)}$ identity matrix.
\end{theorem}

For the sake of brevity, we omit the proofs related to the Lemma 6.1 and the Theorems 6.2 and 6.3, because these exactly follow the procedures exhibited in the Lemma 2.1 and in the Theorems 2.1 and 3.1, respectively. Without any loss of generality, by following the arguments provided at the end of the provious section, we can conclude that we have obtained a general strategy to get a synthetic representation of an arbitrary quantum circuit also in a the general framework of qudits.
We close the section with the following example.

\subsection{An example: the $_{[3]}\sqrt{SWAP}_{(3;1,3)}.$}

Let us suppose to wish to apply the $\sqrt{SWAP}$ to the first and the third qutrits in a three dimensional input state $\ket x \otimes \ket y \otimes \ket z$. 
The expression of the $\sqrt{SWAP}$ gate for two adjacent qutrits has been already studied \cite{GC,W} and it is easy to verify that its block-matrix representation is given by the matrix:
$$_{[3]}\sqrt{SWAP}=\left[ \begin{array}{c|c|c}
					 \sqrt{S_{11}} & \sqrt{S_{12}}   & \sqrt{S_{13}} \\   \hline 
					 \sqrt{S_{21}} & \sqrt{S_{22}}   & \sqrt{S_{23}} \\ \hline
             \sqrt{S_{31}} & \sqrt{S_{32}}   & \sqrt{S_{33}}   \\  
						\end{array}\right],$$

where we have considered to divide $\sqrt{SWAP}$ into the $3^2$ blocks (each one of dimension $3$): 
\begin{eqnarray*}
& \sqrt{S_{11}}=\left[ \begin{array}{ccc}
					 1 & 0  & 0 \\   
					 0 & \frac{1+i}{2}   & 0 \\ 
             0 & 0  & \frac{1+i}{2} \\  
						\end{array}\right], \sqrt{S_{12}}=\left[ \begin{array}{ccc}
					 0 & 0   & 0 \\   
					 \frac{1-i}{2} & 0   & 0\\ 
           0 & 0   & 0  \\  
						\end{array}\right], \sqrt{S_{13}}=\left[ \begin{array}{ccc}
					 0 & 0  & 0 \\  
					 0 & 0   & 0 \\ 
            \frac{1-i}{2} & 0   & 0   \\  
						\end{array}\right], \\ & \sqrt{S_{21}}=\left[ \begin{array}{ccc}
					 0 & \frac{1-i}{2}  & 0\\  
					 0 & 0  & 0 \\
            0 & 0   & 0 \\  
						\end{array}\right], \sqrt{S_{22}}=\left[ \begin{array}{ccc}
					 \frac{1+i}{2} & 0   & 0 \\    
					 0 & 1  & 0\\ 
             0 & 0  & \frac{1+i}{2}  \\  
						\end{array}\right], \sqrt{S_{23}}=\left[ \begin{array}{ccc}
					 0 & 0  & 0 \\  
					 0 & 0  & 0\\ 
             0 & \frac{1-i}{2}  & 0   \\  
						\end{array}\right], \\
& \sqrt{S_{31}}=\left[ \begin{array}{ccc}
					 0 & 0   & \frac{1-i}{2} \\   
					0 &0    & 0\\ 
             0 & 0   & 0 \\  
						\end{array}\right], \sqrt{S_{32}}=\left[ \begin{array}{ccc}
					0 & 0  & 0 \\    
					 0 & 0  & \frac{1-i}{2}\\ 
             0 & 0   & 0 \\  
						\end{array}\right], \sqrt{S_{33}}=\left[ \begin{array}{ccc}
					 \frac{1+i}{2} & 0   & 0\\   
					 0 & \frac{1+i}{2}  & 0 \\ 
            0 & 0 & 1   \\  
						\end{array}\right].
\end{eqnarray*}

By taking into account Theorem \ref{QT}, the required operator is the $3^3$-dimensional square matrix given by:
$$_{[3]}\sqrt{SWAP}_{[3;1,3]}=\left[ \begin{array}{c|c|c}
					 \sqrt{S_{11}}^{(2)} & \sqrt{S_{12}}^{(2)}   & \sqrt{S_{13}}^{(2)} \\   \hline 
					 \sqrt{S_{21}}^{(2)} & \sqrt{S_{22}}^{(2)}   & \sqrt{S_{23}}^{(2)} \\ \hline
             \sqrt{S_{31}}^{(2)} & \sqrt{S_{32}}^{(2)}   & \sqrt{S_{33}}^{(2)}   \\  
						\end{array}\right],$$

where $\sqrt{S_{ij}}^{(2)}=I\otimes\sqrt{S_{ij}}$ (where $I$ is the $3$-dimensional identity matrix).

It is straightforward to check that:
\begin{enumerate}
\item Given three qutrits $\ket x$, $\ket y$, $\ket z$, is $$(_{[3]}\sqrt{SWAP}_{[3;1,3]}\cdot_{[3]}\sqrt{SWAP}_{[3;1,3]})(\ket x \otimes \ket y \otimes \ket z)=(\ket z \otimes \ket y \otimes \ket x);$$
\item $_{[3]}\sqrt{SWAP}_{[3;1,3]}$ is unitary.
\end{enumerate}

\section{Towards a multi-target quantum computational cogic}

In addition to the potential benefits that the representation provided in Section 4 can have in the context of programming language (as discussed in Section 6), this representation can also be considered as a very helpful tool for a further theoretical investigation. This section is devoted to provide an insight about the application of the results obtained in the previous sections on the context of the quantum computational logic (QCL).

The quantum computational theory has naturally inspired new forms of quantum logic, the so called \emph{quantum computational logic} \cite{DGG,DGSL,GLSP,SA}. From a semantic point of view, any formula of the language in the QCL denotes a piece of quantum information, i.e. a density operator living in a complex Hilbert space whose dimension depends on the linguistic complexity of the formula. Similarly, the logical connectives are interpreted as special examples of quantum gates. Accordingly, any formula of a quantum computational language can be regarded as a logical description of a quantum circuit. 
The initial concept at the very background of the QCL is the assignment of the truth value of a quantum state that represents a formula of the language. Conventionally, the QCL assumes        to assign the truth value ``false" to the information stored by the qubit $\ket 0$ and the truth value ``true" to the qubit $\ket 1.$ Unlike the classical logic, QCL turns out to be a \emph{probabilistic} logic, where the qubit $\ket\psi=c_0\ket 0 +c_1\ket 1$ logically represents a ``probabilistic superposition" of the two classical truth values, where the \emph{falsity} has probability $|c_0|^2$ and the \emph{truth} has the probability $|c_1|^2$. As in the qubit case, by the standard approach of the QCL it is also defined a probability function $\texttt p$ that assings a probability value $\texttt p(\rho)$ to any density operator $\rho$. Intuitively, $\texttt p(\rho)$ is the probability that the quantum information stored by $\rho$ corresponds to a \emph{true} information. 

In order to define the function $\texttt p$, we first need to identify in the space $\otimes^n\mathbb C^2$ the two operators $P_0^{(n)}$ and $P_1^{(n)}$ as the two special projectors that represent the \emph{falsity} and the \emph{truth} properties, respectively. Before this, a step is very crucial. In order to extend the definition of \emph{true} and \emph{false} from the space $\mathbb C^2$ of the qubits to the space $\otimes^n\mathbb C^2$ of the tensor product on $n$ qubits (say \emph{quantum register} or, briefly, \emph{quregister}), the standard approach of the QCL accords with the following convention: a quregister $\ket x=\ket{x_1,\dots,x_n}$ is said to be \emph{true} if and only if $x_n=1$; conversly, it is said to be false if and only if $x_n=0$. Hence, the truth value of a quregister only depends on its last component; simply speaking, only the last qubit is considered to assume the role of the target qubit. On this basis, it is natural to define the property  \emph{falsity} (or \emph{truth}) on the space $\otimes^n\mathbb C^2$ as the projector $P_0^{(n)}$ (or $P_1^{(n)}$) onto the span of the set of all \emph{false} (or \emph{true}) registers. Now, accordingly with the Born rule, the probability that the state $\rho$ is \emph{true} is defined as: 
\begin{eqnarray}\texttt p(\rho)=Tr(P_1^{(n)}\rho).
\end{eqnarray}
In the QCL the evolution of a quregister is dictated by the application of a unitary operator  while the evolution of a density operator is dictated by the application of a quantum operation. Of course, for any quantum gate $U$ there exists the correspondent quantum operation $^\mathcal DU$ that replaces the behaviour of the quantum gate in the context of the density operators (in particular, $^\mathcal DU(\rho)=U\rho U^{\dagger}$), but the other way generally does not hold. 

%

Basing on this approach and inspired by the intrinsic properties of the quantum systems, the semantic of the QCL   turns out to be strongly non-compositional and context dependent. This approach, that may appear \emph{prima facie} a little strage,   leads to the benefit to reflect pretty well plenty of informal arguments that are currently used in our rational activity. A  detailed description of the QCL and its algebraic properties are summarized in \cite{DGG}.  
 
Despite its remarkable expressive power, the convention to assume that the target qubit is only the last one forces the QCL to include in the language only one target gates (such as unary gates, C-Not, Toffoli etc...). This restriction is basically unnecessary and it could also seem to be a little far from the architecture of a real quantum circuit. For this reason turns out to be useful to provide an extension of the QCL (that we will call \emph {Multi Target} QCL, briefly MT-QCL) that overcomes this restriction. The immediate benefit of the MT-QCL with respect to the QCL is given by the fact that the first allows to involve in the language also non one target gates (for instance the $Swap$ gate, the $\sqrt{Swap}$ gate and the Fredkin gate) without any lost of generality. Further, in this framework the standard QCL can be seen as a particular (one target) instance of the MT-QCL.

Similarly to the QCL case, the essencial step in the introduction of the MT-QCL is the definition of probability. Let us consider a simple computational system given by the $n$-dimensional input state $\ket x= \ket{x_1,\dots,x_n}$ and one operator $U^{(n)}$ acting on the space $\otimes^n\mathbb C^2$ as $U^{(n)}\ket{x}=\ket{y_1,\dots,y_n}$. Let us consider the two following sets of indexes dictated by $U^{(n)}$: $$C_{U^{(n)}}=\{i:\ket{x_i}=\ket{y_i}\} \,\,\, and \,\,\, T_{U^{(n)}}=\{j:\ket{x_j}\neq\ket{y_j}\}.$$

Intuitively, $C_{U^{(n)}}$ selects the position of the qubits of the input that are affected by $U^{(n)}$; conversely for $T_U^{(n)}.$ Conveniently, let us call any $i$ belonging to $C_{U^{(n)}}$ a \emph{control position} and any $j$ belonging to $T_{U^{(n)}}$ a \emph{target  position}.\footnote{Let us give a slight abuse of the terms \emph{target} and \emph{control} according with the convention that the \emph{control position} is related to the qubit that is not affected by the gate, otherwise we speak about \emph{target position}.}
On this basis, we define a probability $\texttt P$ associated to the couple $[U^{(n)},\ket x]$ as: 

\begin{definition}\label{Prob}  

$$\texttt P[U^{(n)},\ket x]=Tr(\mathcal P_1\otimes\cdots\otimes\mathcal P_n)^{\mathcal D}U^{(n)}\rho_{\ket x}$$
where $U^{(n)}$ ia a $n$-dimensional unitary operator, $\rho_{\ket x}=\ket x\langle x|$ and

$\mathcal P_i=
I, \,\,\,\, if \,\,  i\in C_{(U^{(n)})}$ and\,
$\mathcal P_i=P_1, \,\,\,\, if \,\,  i\in T_{(U^{(n)})}.$


\end{definition}

The definition can be naturally generalized, without any lost of generality, to the case where the input state $\rho$ is a mixed state (in this case we write $\texttt P[^{\mathcal D}U^{(n)},\rho]$). 
At this stage, the natural continuation of the investigation should be devoted to: $i)$ study the behavior of this probability applied to coulpes given by one target/non one target gates and product/non product input states; $ii)$ make a full comparison between the QCL and the MT-QCL; $iii)$ show the semantic advantages provided by the MT-QCL exploiting the possibility to dispose of a larger language with respect to the QCL. But, for the aims of this paper, we confine to provide, as an example, the probability value of an arbitrary binary gate applied to an arbitrary (non product) state, showing how the block-matrix  representation given by the Theorem (\ref{U}) plays a crucial role.

\begin{theorem}\label{binary2}
\nl

Let $U$ a binary operator represented as in Section (\ref{4}): $U=\left[ \begin{array}{c|c}
					U_{11} & U_{12} \\ \hline     
					 U_{21} & U_{22} \\ 
						\end{array}\right]$   (let $U$ be not a control-target gate) and let $\rho\in\otimes^k\mathcal D(\mathbb C^2).$ Then:
$$\texttt P[U_{[k;m,m+n]},\rho]=Tr[(\Lambda_U^{(m+n)}\otimes I^{(k-m-n)})\rho],$$ where $\Lambda_U^{(n+1)}=\left[ \begin{array}{c|c}
					I^{(n-1)}\otimes(U_{21}^{\dagger}P_1U_{21}) & I^{(n-1)}\otimes(U_{21}^{\dagger}P_1U_{22})  \\ \hline     
					 I^{(n-1)}\otimes(U_{22}^{\dagger}P_1U_{21}) & I^{(n-1)}\otimes(U_{22}^{\dagger}P_1U_{22})  \\ 
						\end{array}\right].$
\nl

Proof:
\begin{eqnarray*}
& & \texttt P[U_{(k;m,m+n)},\rho]=\\
&=& Tr[(P_1^{(m)}\otimes P_1^{(n)}\otimes I^{(k-m-n)})U_{(k;m,m+n)}\,\,\rho\,\, U_{(k;m,m+n)}^{\dagger}] =\\
&=& Tr[(U_{(k;m,m+n)}^{\dagger}(P_1^{(m)}\otimes P_1^{(n)}\otimes I^{(k-m-n)})U_{(k;m,m+n)})\,\rho]=\\
 & & (by\,\, Theorem\, (\ref{U}))
= Tr[(I^{(m-1)}\otimes  \left[ \begin{array}{c|c}
					U_{11}^{(n)\dagger} & U_{21}^{(n)\dagger} \\ \hline     
					 U_{12}^{(n)\dagger} & U_{22}^{(n)\dagger} \\ 
						\end{array}\right] \otimes I^{(k-m-n)})\cdot \\ 
&\cdot& (I^{(m-1)}\otimes (P_1\otimes P_1^{(n)})  \otimes I^{(k-m-n)}) \cdot \\
&\cdot& (I^{(m-1)}\otimes \left[ \begin{array}{c|c}
					U_{11}^{(n)} & U_{12}^{(n)} \\ \hline     
					 U_{21}^{(n)} & U_{22}^{(n)} \\ 
						\end{array}\right]  \otimes I^{(k-m-n)}) \rho]=\\
&=& Tr[(I^{(m-1)}\otimes(\left[ \begin{array}{c|c}
					U_{11}^{(n)\dagger} & U_{21}^{(n)\dagger} \\ \hline     
					 U_{12}^{(n)\dagger} & U_{22}^{(n)\dagger} \\ 
						\end{array}\right]\cdot \left[ \begin{array}{c|c}
					0 & 0 \\ \hline     
					 0 & P_1^{(n)} \\ 
						\end{array}\right]\cdot \left[ \begin{array}{c|c}
					U_{11}^{(n)} & U_{12}^{(n)} \\ \hline     
					 U_{21}^{(n)} & U_{22}^{(n)} \\ 
\end{array}\right])\otimes I^{(k-m-n)})\,\rho].
\end{eqnarray*}
Let us notice that
\begin{eqnarray*}
&  \left[ \begin{array}{c|c}
					U_{11}^{(n)\dagger} & U_{21}^{(n)\dagger} \\ \hline     
					 U_{12}^{(n)\dagger} & U_{22}^{(n)\dagger} \\ 
						\end{array}\right]\cdot \left[ \begin{array}{c|c}
					0 & 0 \\ \hline     
					 0 & P_1^{(n)} \\ 
						\end{array}\right]\cdot \left[ \begin{array}{c|c}
					U_{11}^{(n)} & U_{12}^{(n)} \\ \hline     
					 U_{21}^{(n)} & U_{22}^{(n)} \\ 
\end{array}\right]=\\
&= \left[ \begin{array}{c|c}
I^{(n-1)}\otimes(U_{21}^{{\dagger}}P_1U_{21}) & I^{(n-1)}\otimes(U_{21}^{{\dagger}}P_1U_{22}) \\ \hline      I^{(n-1)}\otimes(U_{22}^{{\dagger}}P_1U_{21}^{(n)}) & I^{(n-1)}\otimes(U_{22}^{{\dagger}}P_1U_{22}) \\ 
\end{array}\right]=\Lambda_{U}^{(n+1)}
\end{eqnarray*}
Hence, $$\texttt P[U_{(k;m,m+n)},\rho]=Tr[(I^{(m-1)}\otimes\Lambda_U^{(n+1)}\otimes I^{(k-m-n)})\rho]=Tr[(\Lambda_U^{(m+n)}\otimes I^{(k-m-n)})\rho].$$ 

\end{theorem}

This Theorem allows to easily obtain the probability value of an arbitrary binary gate applied to an arbitrary input state. A generalization of this result to $n$-ary gates, including in the framework also qudits, and a complete investigation on the semantic structure of the MT-QCL will lead to obtain a very new and general model of quantum computational logic, as will be fully developed in a future work. As showed in the previous Theorem, the block-matrix representation provided in this paper turns out to be essential for the achievement of these results and a complete development of the MT-QCL will repeatedly require the utilization of this representation.

\section{Conclusions and further developments}
The main purpose of this work is to provide a kind of simplification of the language of the quantum circuits, by exploiting the block-matrix representation of arbitrary quantum gates.  We have shown a strategy that allows to represent an arbitrary quantum gate (and, in principle, a sequence of quantum gates) applied to arbitrary qubits of the input state, without incurring in the necessity to consider the composition of multiple binary \emph{SWAP} gates among these qubits. Indeed, even if it represents a very common scenario in the architecture of quantum computation, a systematic mathematical representation of this picture was actually missing.  We have also provided a generalization of this model where the input state is given by a composition of qudits. 

This model represents a mathematical tool that could be exploited, in principle, for all the computational problems related to the architecture in quantum computer design in order to suggest suitable strategies able to lead to concrete computational benefits. On the other hand, an immediate utilization of this representation can be conducted in the context of the simulation of the quantum circuit by using classical programming languages, in order to obtain very flexible packages to represent complex quantum circuits through a standard classical computer.

A further theoretical development can be performed in the context of the quantum computational logic, where the target bit is generally assumed to be only the last qubit (or qudit) \cite{DGSL} within a given quantum circuit. Indeed, in a more realistic scenario, the target bit has not to be unique and it could occupy an arbitrary position over a quantum circuit. On this basis, the results provided in this work suggest a generalization of the language of the quantum computational logic where multiple target qubits placed in arbitrary positions are considered,  in order to define a kind of multiple-target quantum computational logic. The last section of this paper is devoted to provide an insight of this idea and to show the utility of the block-matrix representation also for this purpose. A full description of a multi target quantum computation logic and a complete investigation on its logical and algebraic properties is hereby proposed as a further development. 

\bigskip

\section*{Acknowledgements}
This work is partially supported by Regione Autonoma della Sardegna within the project ``\emph{Time-logical evolution of correlated microscopic systems}"; CRP 55, L.R. 7/2007 (2016).
I also thank some of the Reviewers for the insightful remarks and  Dr. Federico Holik for the useful suggestions in the revision of the paper.

\bigskip





\end{document}